# DevOps Capabilities, Practices, and Challenges: Insights from a Case Study

Mali Senapathi
Auckland University of Technology
Auckland, New Zealand
mali.senapathi@aut.ac.nz

Jim Buchan
Auckland University of Technology
Auckland, New Zealand
jim.buchan@aut.ac.nz

Hady Osman
Auckland, New Zealand
hadyos@gmail.com

## ABSTRACT

DevOps is a set of principles and practices to improve collaboration between development and IT Operations. Against the backdrop of the growing adoption of DevOps in a variety of software development domains, this paper describes empirical research into factors influencing its implementation. It presents findings of an in-depth exploratory case study that explored DevOps implementation in a New Zealand product development organisation. The study involved interviewing six experienced software engineers who continuously monitored and reflected on the gradual implementation of DevOps principles and practices. For this case study the use of DevOps practices led to significant benefits, including increase in deployment frequency from about 30 releases a month to an average of 120 releases per month, as well as improved natural communication and collaboration between IT development and operations personnel. We found that the support of a number of technological enablers, such as implementing an automation pipeline and cross functional organisational structures, were critical to delivering the expected benefits of DevOps.

## 1 INTRODUCTION

The DevOps concept [1] emerged to bridge the disconnect between the development of software and the deployment of that software into production within large software companies [2]. The main purpose of DevOps is to employ continuous software development processes such as continuous delivery, continuous deployment, and microservices to support an agile software development lifecycle. Other trends in this context are that software is increasingly delivered through the internet, either server-side (e.g. Software-as-a-Service) or as a channel to deliver directly to the customer, and the increasingly pervasive mobile platforms and technologies on which this software runs [3]. These emerging trends support fast and short delivery cycles of delivering software in the fast-paced dynamic world of the Internet. As such DevOps has been well received in the software engineering community and has received significant attention particularly in the practitioner literature [4]. Annual 'State of DevOps' reports show that the number of DevOps teams has increased from 19% in 2015 to 22% in 2016 to 27% in 2017 [5].

However, as observed in recent studies, despite their growing popularity, there is a lack of empirical research on the actual practice of DevOps beyond a discussion of blog posts and industrial surveys [6, 7]. Beyond very few case studies [8], the current literature does not provide much insight on the actual implementation and practices of DevOps and their effectiveness in supporting continuous software development. In this research, we investigate these issues based on an in-depth exploratory case study. In particular, we aim to address the following research questions:

- what are the main drivers for adopting DevOps?
- what are the engineering capabilities and technological enablers of DevOps?
- what are the benefits and challenges of using DevOps?

## 2 RELATED RESEARCH

The concept of DevOps has been described as ambiguous and difficult to define [7]. While there is no standard definition for DevOps, two main opposing views exist in the blogosphere [6, 7, 9]. One view identifies DevOps as a specific job description that requires a combination of software development and IT operations skills, and the other argues that the spirit of DevOps addresses an emerging need in contemporary software development rather than a job position. In an attempt to address this issue, one of the two main streams of research in DevOps has strived on achieving a clear understanding of (i) of definitions and characterization of DevOps and its associated practices [7, 10-13], and (ii) the benefits and challenges of adopting DevOps [7, 8]. For example, while Culture, Automation, Measurement, Sharing, Services have been identified as the main dimensions of DevOps [10], others have described it as a cultural movement that enables rapid development with four defining characteristics: open communication, incentive and responsibility alignment, respect, and trust [14]. The significance of cultural change in improving the collaboration between development and operations in order to accelerate delivery of changes is stressed [11]. On the contrary, it has been argued that cultural aspects by themselves cannot be the defining characteristics of DevOps, but rather act as enablers to support a set of engineering process capabilities [7].

The second stream of research focuses on understanding the challenges and benefits associated with adopting practices such as continuous delivery and continuous deployment, which serve as the basic building blocks of a working agile/DevOps implementation [4]. This includes growing number of empirical studies discussing benefits and challenges of continuous integration [15, 16], continuous delivery [17, 18], and continuous deployment [19, 20]. Fitzgerald and Stol [3] label all these continuous activities together as 'Continuous *' (i.e. Continuous Star) practices and highlight the need for a more holistic and integrated approach across all the activities that comprise software development. According to Dingsøyr & Lassenius [4], all these emerging topics, i.e. DevOps and continuous practices come under the umbrella of continuous value delivery.

In summary, while the first stream of research has largely centered on understanding the conceptual and defining characteristics of DevOps, the second stream has focused on understanding the benefits and challenges of adopting some of the 'Continuous *' practices and argues for an increased interest in these emerging topics. Little is known about how DevOps is actually implemented in real software development practice. Therefore, it is especially pertinent to understand the use of DevOps in a real product development setting, where experienced software developers adopted a gradual and customised approach to its implementation. We believe that the lessons learned from its implementation in a real software development context are invaluable, as few such studies have been published.

Given the above, we used the DevOps definition developed by [7] as a guiding framework to investigate the implementation of DevOps in actual practice.

## 3 RESEARCH FRAMEWORK

The following definition encapsulates many of the ideas and concepts identified by other authors, and added a useful structure to describe and analyse DevOps and its enablers:

"a set of engineering process capabilities supported by cultural and technological enablers. Capabilities define processes that an organisation should be able to carry out, while the enablers allow a fluent, flexible, and efficient way of working" [7].

The three core aspects in this definition are DevOps capability enablers, cultural enablers, and technological enablers. Table 1 lists the technological and capability enablers, the focus of this paper. In [7] The cultural and technological enablers are viewed as supporting the capability enablers.

**Table 1 Enablers of DevOps (adapted from Smed & colleagues [7]):**

| Capabilities | Collaborative and continuous development<br>Continuous integration and testing<br>Continuous release and deployment<br>Continuous infrastructure monitoring and optimization<br>Continuous user behavior monitoring and feedback<br>Service failure recovery without delay<br>Continuous Measurement |
|---|---|
| Technological Enablers | Build automation<br>Test automation<br>Deployment automation<br>Monitoring automation<br>Recovery automation<br>Infrastructure automation<br>Configuration management for code and infrastructure<br>Metrics automation |

The DevOps capability enablers incorporate the basic activities of software development (i.e. planning, development, testing, and deployment) carried out continuously based on feedback from other activities. For example, the continuous deployment capability facilitates deployment of new features a soon as they have been integrated and tested successfully. This, however, requires the support of technical practices such as test automation and effective collaboration between the development and deployment teams. The feedback data on service infrastructure performance, as well as how and when the users interact with the service, is encapsulated by the two capabilities of infrastructure monitoring and user behavior monitoring. These capabilities provide valuable input to the planning and development processes to improve and optimize the service. Finally, a DevOps organisation should have the necessary monitoring infrastructure to detect service failures and the capability to recover from such failures immediately.

The technological enablers support the DevOps capabilities by automating tasks. Automation facilitates continuous delivery and deployment by providing a single path to production for all changes to a given system, whether to code, infrastructure and configuration management environments [21], where custom programs or scripts configure and monitor the service infrastructure. The cultural enablers relate to behaviours that DevOps teams must exhibit in order to support the DevOps capabilities in a positive way. They emphasise the need for extensive collaboration and low effort communication, shared goals, continuous experimentation and learning, and collective ownership.

We have added two enablers to the original framework by Smeds and colleagues [7], related to metrics. We argue that collecting empirical evidence of achieving (or not) DevOps-related goals is an important driver for deciding whether to make changes (or not) to the DevOps implementation.

Technologies and team capability to measure improvements towards goals are enablers of DevOps evolution. Automation of metric measurement is a technological enabler of DevOps in the sense it can support the team's capability of continuous measurement of appropriate metrics. The metrics automation may be implemented through specific tools, or through instrumentation of existing tools. Which metrics are important to continuously measure through automation will be context dependent.

## 4 BACKGROUND TO THE CASE

The case organisation is a New Zealand-based software company in the Finance/Insurance sector that delivers services for small and medium-sized businesses through a cloud-based software product suite developed in-house. The company is high growth and has offices in New Zealand, Australia, the United Kingdom, the United States and Singapore. Its products are based on the software as a service (SaaS) model and sold by subscription. Its products are used in over 180 different countries.

The software development process is based on Agile values and principles and implemented through Scrum practices and roles in general. The teams have 2 to 3 week sprints that include daily stand-up meetings, sprint planning and sprint review meetings, and sprint retrospectives.

The development teams are cross-functional, self-organizing and organised by product functional module. The roles in the development teams vary from team to team but typically include Developers, Testers, a Product Owner and an Agile Facilitator, with shared support from members of the wider product team.

The company examined in our study was around one year into DevOps adoption, after establishing the need for a change by the business in order to remain agile and competitive. Prior to DevOps implementation the company's product team was split into two separate delineated teams: platform and product development, with the former having exclusive access to production systems. Prior to DevOps, the company had been maintaining and developing its aging monolith application that was hosted in a traditional data center. While this model was able to serve the company well and contribute to its success of shipping software quickly in its early stages, it had numerous shortcomings that quickly became visible to the business. As a result, the company undertook a number of fundamental changes. Early on, they commissioned a costly migration of hosting providers to one that provided on-demand cloud computing platform. This change allowed product teams to access and maintain their own independent infrastructure, and gave them autonomy to work much closer with engineers to design and build what they needed providing end-to-end control. A big part of the expense of this exercise was spent in rewriting large parts of their monolith application to work in this new platform environment that scaled independently and had different uptime Service Level Agreements than before.

From a team perspective, the company introduced an "embedded operations model" by disbanding the silo of the operations team and moving platform engineers into product development teams. Aside from their existing duties, the product development teams then became responsible for operations and cost of their own platform with their newly acquired operations skillset. The focus was on creating cross-functional teams that had end-to-end capability and incentives for shipping product and operating it. The creation of such teams involved investing in acquiring the right skill set.

A number of centralised platform functions (security, data services, shared components, etc.) were still retained by the company, however, they were now acting as service providers to their new internal customer, the product development team.

## 5 RESEARCH METHODOLOGY

We adopted a case study methodology as it enables investigation of a contemporary phenomenon within its natural setting and is appropriate for contemporary topics such as DevOps where theory and practice are relatively new [23].

Data collection involved a series of six in-depth semi-structured one-on-one interviews, conducted over a six-month period with interviewees covering the spectrum of the key roles responsible for DevOps implementation, namely: Developer (Dev), Tester (T), Release Quality Lead (RQL), Team Lead Infrastructure (TLI), Training Manager (TM), and Operations Manager (OM). Interviews were generally of 1-1.5 hour duration, and were followed up by some informal sessions to clarify and refine issues as they emerged. Smeds's [7] model was used to develop an interview protocol. Interviews allowed the researchers to explore the interviewee's view of the DevOps implementation process, particularly the main drivers, engineering capabilities and technological enablers, benefits and challenges associated with adopting DevOps. The responses of the interviewees included information on multiple projects. All interviews were digitally recorded with the permission of the participants and later transcribed in detail.

The transcribed data were uploaded into the qualitative analysis tool NVivo. Individual interview transcripts were analysed for concepts or themes by one researcher. The coded themes were re-analysed to ensure that they belonged to the correct category. This continued until the conceptual categorisation we developed was well-supported by the data.

In order to clarify some details about the pre-DevOps situation in the organisation and clarify some of the drivers with the initiators of the DevOps adoption, one of the authors had a short post-interview conversation with the pre-DevOps Chief Product Officer and Chief Platform Officer. The outcome of this discussion provided a better understanding of the main drivers that motivated the adoption of DevOps in the case organization. However, it was not included when analysing the interview data.

## 6 FINDINGS

Our findings from the analysis of the interview transcripts are discussed in the following sections. First, we present an overview of the DevOps journey from perspective of the main concepts and definitions associated with the meaning of DevOps in the organisation. This is followed by a description of the organisation's main drivers and motivation for adopting DevOps (i.e. the expected benefits). The technological and capability enablers of the organisation's DevOps implementation are then examined, followed by a discussion of the benefits of DevOps that the organization has realized so far. We finish the findings with an analysis of the challenges that hindered the effective implementation of DevOps.

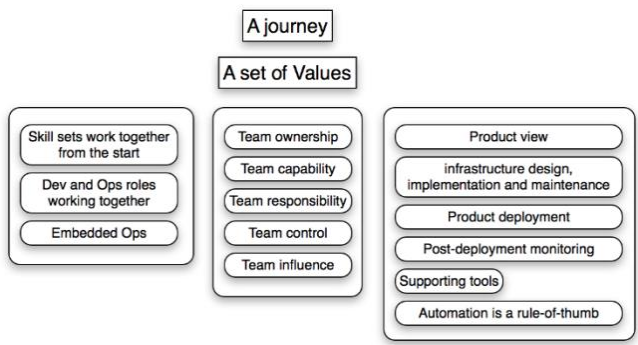

**Figure 1: The Meaning of DevOps**

### 6.1 The Meaning of DevOps

Interviewees offered a number of interesting perspectives on the meaning and conceptualization of DevOps, having experienced its adoption for around a year. The main concepts are depicted in Figure 1.

At a high level, DevOps was viewed as a journey and a set of values that guided behavior. It was recognized that DevOps adoption was incremental and a transitional journey. For example, the TM describes DevOps as a "*period of time where software developers transition from just handing over their completed work to system administrators, to actually taking ownership and responsibility themselves*".

A technical value described by the TL was "*looking at automation as a rule of thumb*".

DevOps was commonly described as a way of bringing the skills and knowledge of operations and development closer, with greater collaboration and communication.as the RQM describes it as "*…a kind of hands-on, short-term and longer-term situation where everybody's working really closely, communicating really closely, and getting an understanding of where everything's at so they're not just two very segregated departments [anymore]".* The OM emphasized that DevOps is about "*more natural communications with the people around you*". The term "embedded Ops" has been adopted in the organisation to describe the situation where Product teams have a dedicated Ops specialist as part of the team. At the time of the interviews not all development teams had transitioned to this situation.

Product team members tended to view DevOps from the perspective of an end-to-end product view with broader team responsibilities and control. As the tester described it, "*We write stuff, we review it, we test it, and we deploy it. And through that as well as discussion and learning, and that kind of thing. It's pretty choice*". The Dev also emphasizes this "team control" view of DevOps: *"you're not relying on other teams to do the infrastructure. So, you manage your own infrastructure. You have control over it."*

He goes on to explain his view of the impact of this autonomy: "*If you wanted to use a specific tooling you can…and as a Dev it's a lot easier to code if that's the right tool for the job and it's a lot easier to deploy and everything…. But because you take care of the environment you are in charge of the cost and taking care of it. So, you do consciously think about it [more]".* Interviewees also noted that the team's understanding of DevOps included responsibility for writing infrastructure scripts as well as ownership of post-deployment monitoring of infrastructure and issue resolution.

The OM describes his view of DevOps space by tracing it back to the roots of computer engineering:

".. *you have to have quite creative mind-set, have this weird like sort of spatial, cognitive space between science and arts that comes into play in the way you build these computer systems. A lot of it's kind of Lego bricks. You can just put things together and you can see how they interact, and a lot of it's reusable stuff and that's sort of how you think about building out an entire environment and a product. Because at the end of the day it's really becoming like product is the shell and everything else goes into it to support it.*"

## 6.2 Drivers for DevOps Adoption

Transforming a traditional product organisation to adopt a DevOps model can be both an expensive and time-consuming undertaking. Yet many rapidly growing organisations justify investment in this transformation because the expected benefits accrued from the outcomes are greater than the cost of effort and change to undertake the DevOps implementation journey. The expected benefits, or drivers, that motivate DevOps adoption for the case organisation are depicted graphically in Figure 2 including strategic, tactical and operational drivers.

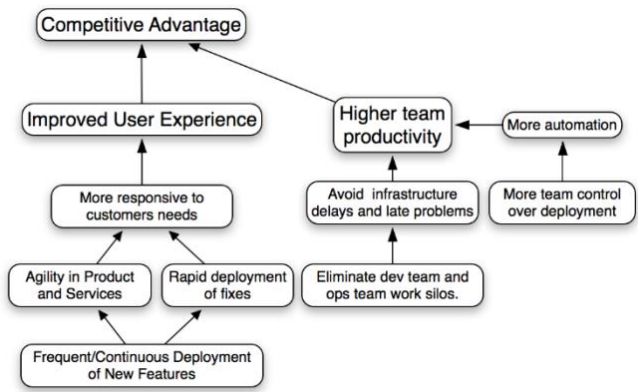

**Figure 2: Drivers for implementing DevOps**

Firstly, a strategic view is provided by a short post interview discussion with the pre-DevOps Chief Product Officer and Chief Platform Officer. They describe three pre-DevOps frustrations that motivated the adoption of DevOps and initiated the work to move away from a centralised operational model. Firstly, was the frequent frustration between the company's operation and product teams who have had competing priorities because of a "*separation in the wrong part of the value chain. Product teams are required to ship product quickly, often with networking and operational changes needed. Operation teams serve requests from many multiple teams and set their own internal priority without often taking into account product team timelines. Working as silos naturally created points of frustration because of lack of alignment between the two units*".

Secondly, the Operation and Product teams operated under what was identified as a mismatch of incentives and control. Operation teams were accountable for performance and uptime, yet development teams were in a better position to improve it. Conversely, development teams were accountable for shipping product with great agility and velocity, but operation teams were in controls of major portions of the software development lifecycle (SDLC).

Lastly, as the organisation utilized more technological enablers and in particular automation for more agility, the need to move to a hosting provider that allowed for infrastructure as code also grew. This move required a different skill set that is more aligned to developers in development teams.

The driver for DevOps adoption most emphasized by interviewees was to achieve continuous deployment (CD), "*the ability to be able to make a change and have that reflected in the real world, instantly...*" (ITL). As depicted in Figure 2 this driver relates to more strategic expected benefits including a higher responsiveness to customers, through faster new feature delivery and bug fixing. CD also "*avoid[s] the outages needed for large releases*" [OM]. SO, changing the pre-DevOps situation, where new product versions were released several times a year, to continuous deployment, was viewed as a strong strategic driver for adopting DevOps.

Another key (tactical) driver for DevOps adoption in the organisation was to achieve productivity improvements or "*deliver[ing] quality software at speed*" [TM]. As seen in Figure 2, this driver relates to other operational drivers. For the OM and TM, getting the Infrastructure Team and Development teams out of their work silos and working more closely together was a strong driver for DevOps adoption. In the pre-DevOps situation "*there was a bottleneck to get stuff into production because we had to give it to the Ops team*" [TM]. The Infrastructure Team would only understand the infrastructure needs and put it in place and deploy *after* the commit. "*.. being able to deliver quality software quickly, you need to have less points along the journey*" [TM], and DevOps realized this. Avoiding "*the double ups and start-stops in communications between ops and devs dealing with an issue ticket*" [OM] was also an expected benefit related to elimination of work silos from DevOps adoption.

From the Development Team's perspective a key (operational) driver for DevOps adoption was "*for the production team to own the infrastructure*" [T]. The Developer's perspective has an interesting perceived benefit: "*It just means you are not relying on other teams to do the infrastructure. You have control over it – choice of tool to use for example. To get the feel of small startups in a big organisation*" [Dev]. The Development team were also motivated by the opportunity DevOps adoption provided to automate more of the testing and infrastructure setup.

## 6.3 Technology and Capability Enablers

Enablers are contextual factors that support an effective implementation of the DevOps way of working. Following the research framework described in section 3, Figure 3 represent snapshots of the organisation's current state of technology support and team capability support for implementing DevOps.

The (H), (M) and (L) beside each enabler indicate the level of maturity of the areas of technical support and level of team capability in each area. As can be seen from this, generally the technology is in place to support the implementation of DevOps to a high degree of maturity. Figure 3 shows there are no big gaps in team capability enablers either, apart from continuous measurement.

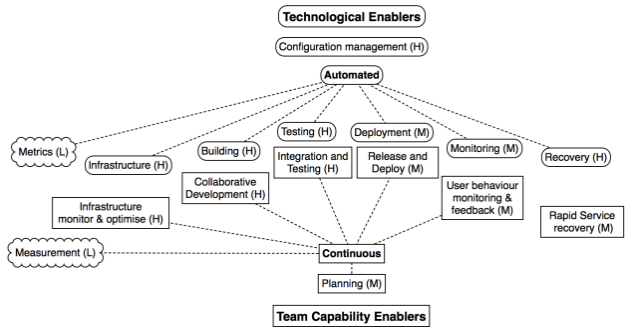

**Figure 3 Technological and team capability enablers**

The following sub-sections provide more detail of the situation with regard to these DevOps enablers. The first sub-section describes the team process capabilities and tool technology support related to aspects of the CI/CD pipeline, with more detail on test automation in the following sub-section. This covers most of the enablers apart from those related to monitoring, which are discussed next. This covers aspects of continuous infrastructure monitoring and optimization and continuous user behavior monitoring and feedback, as well as service failure recovery without delay. The final sub-section discusses the metrics used as evidence of improvement as a result of DevOps adoption.

*6.3.1 CI/CD Pipeline*. For the case organisation, the main goal in implementing DevOps was to achieve continuous delivery and implement the CI/CD pipeline by automating steps in the software delivery process from commit to deploy. Figure 4 summarises the state of the continuous delivery pipeline at the time of this study.

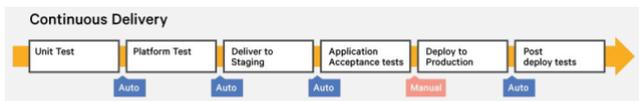

**Figure 4 The CD/CI pipeline**

Continuous delivery was enabled by implementing a set of processes and supporting tools such GoCD, TeamCity, Terraform, and Octopus Deploy. While GitHub was used companywide as a code repository for both product and infrastructure, and quality control around any product or infrastructural changes, Terraform was primarily used for building infrastructure efficiently. TeamCity was used for continuous integration and Octopus Deploy to deploy specific release/version numbers, *"… you create a release in that you pick what you're releasing, like which version numbers… it's a set process that each release must go to. So, you create the release and you want release version number 123. So, if you click "next" on that step, it will roll it to set branch environment that you've configured for it. At that point you know you can kick off testing on that…so, they could be auto tests, or manual tests…then, it might go to the next environment, then it goes live."* [RQL]

Collaborative technologies such as Yammer, FlowDock, and Confluence were used to foster team collaboration. While Flowdock was mainly used for team communication (e.g. keeping in touch, sharing issues/pain points), Yammer was used to share releases with others and to initiate discussion on completed tasks and lessons learnt. Release plans and documentation were stored in Confluence and Jira was used as an issue tracking system to log and track issues such as those relating to building a new piece of software or customer experience.

*6.3.2 Monitoring.* Basic services such as dashboards were used to show information about all releases so that everyone could see in real time mode what was going out. Companywide dashboards showed details such as the total number of users on the system and the countries they come from. There was at least one dashboard associated with every team to look at the infrastructure that supported that area, and as part of taking on their self-deploy the teams had to create dashboards so that they could monitor their piece of the application. This enabled the teams to report on any changes made and customer experience.

Monitoring services such as Datadog and Datawatch were used to monitor metrics such as concurrent user sessions, database load, and CPU metrics. Most teams set up their own Flowdock and set up a link which fed back all the alerting from Datadog into their Flowdock where they could chat real-time on things such as their next release. New Relic was used as a dedicated tool for performance monitoring.

Feature flags were used mainly to control operational aspects from an infrastructure perspective, for example, decisions on resources were made by looking at changes over time by comparing current data with previous trends. Operations feature flags were also used to monitor unclear performance implications of query-time executions such as:

*"..what is the expected behaviour of this app? Is it 60 seconds for a query? Is it going to be longer than that? and if we get that kind of understanding by app, by feature, we can start building some really focused monitoring and automation around that. So, we can start responding to those thresholds in ways that will keep things running smoothly. …".* (Dev)

Monitoring user behavior, although its importance was recognized, was still not very prevalent, as the Tester explained: "Currently not very much but for some specific features, like newly developed features, we do think about monitoring before we develop or when we are developing. Like once the feature is in production, users start using… what stats might be helpful for us to determine whether the feature should have more improvement or it's already good enough or there is something we haven't thought about…"

*6.3.3 Test Automation.* While there were different layers of test automation, most end-to-end functional testing was automated, "*..I think the percentage might be 40 percent for our most used features and for our most common functions we do have auto tests...Unit tests, mostly it's developers. Once they finish a feature, they will develop unit tests for what's added. Once it's deployed to our test environment, it's available for QA to pick up. QA will decide... because from the planning, if we think it's a good candidate for automation, we will create the auto test for this feature, like when they are still developing..*"[T].

In terms of full stack end-to-end testing, developers were involved in doing automated unit testing, whereas mock integration tests were done in test environment, a replica of production where all the integration testing and automated testing would be run, "*..because everything is micro serviced and API-driven we've mocked up API endpoints to test against. So that allows our test environments to be completely isolated from the rest of the company so we can make sure that we have code integrity and no hidden dependencies...and then in our UAT environments we do proper integration tests and acceptance testing*." [OM]. Tools such as Cucumber and Selenium were used to write the tests. Terraform and AWS Cloudformation were used to test Infrastructure as Code, and Selenium for acceptance testing. According to the operations manager, managing infrastructure as code via source control was the philosophy underlying everything that relates to pioneering the DevOps space.

*6.3.4 DevOps Metrics.* At the time of the interviews the organisation had not started systematically collecting metrics, although the need to track improvements in mean time to recover and lead time were mentioned. All interviewees focused on the significant improvements in deployment frequency. For example, teams started realizing that some apps which were deployed fortnightly due to restrictions between dependencies between their apps, "*..that dependency didn't really exist or when it did exist it could be easy avoided. And what they ended up doing was they split all the three things out separately and we could essentially deploy that same app as many times as we wanted it at. I think at one point we even did seven deployments one week which was quite a big deal….*" (Dev)

*6.3.5 Product Architecture.* Several of the interviewees discussed the decision to move to a cloud-based micro-services architecture as an enabler of the DevOps adoption. The ability to reduce dependencies between features as micro-services was seen as a key enabler of fast feature deployment.

## 6.4 Benefits Realised

The drivers or expected benefits of adopting DevOps have been described in section 6.2. Now we describe the benefits actually realized from the DevOps implementation to date, identified by interviewees. The findings are summarized in Figure 5 and discussed in more detail in the following subsections.

*6.4.1 Teams are happier and more engaged.* Although not identified as a driver, this benefit was a strong theme of the interviewees. As shown in Figure 5, there are a number of other DevOps-related benefits that have contributed to the improved team happiness and engagement. Product teams felt more valued in the new DevOps way of functioning. The embedded ops did not feel that they were just sitting in the dark maintaining servers and databases, but could see the value and impact of their work on real clients. DevOps enabled the development team to have a more comprehensive view of the entire landscape, the company, the product and how it is used by clients. As the Operations manager explained, "*You understand how everything fits together; you understand how it works; you actually build your own solutions for things that work for your environment, and not trying to sort of bend an enterprise-type software to suit your whims*"

Interviewees also described how the increased collaboration with others needed to implement DevOps was enjoyable and motivating.

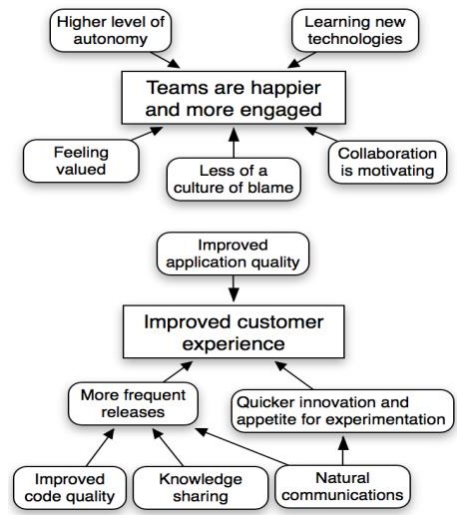

**Figure 5 Benefits realised from DevOps adoption**

Related to this is the decrease in finger pointing in the teams that was reported by interviewees. This was described as contributing to a more positive collaborative team environment. Many of the team members clearly enjoyed learning about new technologies and were motivated by the need to learn about the new DevOps technology enablers as part of their work. The increased responsibilities of the team to include Ops functions was viewed as a benefit by providing more team autonomy in their work. "*Team ownership and responsibility is huge, the Devs and QAs have loved it…*" [RQM]. The TLA viewed this autonomy as enabling the team to "*..build so much better integrity. You build your own solutions that work for your own [team] environment*".

*6.4.2 More frequent releases*. This DevOps drivers was front-of-mind for most interviewees and similarly it was a strong theme as a realized benefit. The benefits accrued from smaller more frequent releases is described by the RQM: "*More frequent releases [is a benefit]. Because [there are] more deployers and smaller releases. Easier to contain a release. More features for end users*". The TM also observed that the smaller more frequent releases were less risky and resulted in fewer service outages.

Shared technical knowledge between operations and development teams is viewed as a benefit from DevOps adoption that contributed to more frequent releases. It helped in diagnosing and fixing problems faster. "..*even if my focus is testing, it still helps a lot if I know that Ops and Development knowledge, technical knowledge. It directly or indirectly affects my testing job. If I know that I can do it more efficiently and more easily. If you see a customer reported a ticket and if it comes to me, if I don't have any knowledge, I will go and find someone else to fix the problem but if I already know development knowledge at least I can do an initial investigation, right?*" [T].

In DevOps, development teams become a part of taking ownership of the production environment, gaining an understanding of infrastructure and the impact of their code, and better application and code quality were benefits identified as a result of this. The Dev's reasoning was "*that you write better code because you know what's going to happen to it*". The RQM explained: "*…the more understanding that the Devs and the QAs have over the infrastructure itself, they can write that quality code, and a better, kind of smarter, code as well….and so, by the teams getting more of an understanding as to how that worked, they actually changed the way they wrote the code*". Before adopting DevOps, the operations personnel were traditional system administrators who looked after the servers and infrastructure without any feedback back to the product teams unless something went fundamentally wrong. By moving from traditional to a cloud hosting platforms, operations could see the power of being able to do automation and configuration management. The operations people also started understanding why the code was written in a certain way, which helped them to design better infrastructure solutions.

Having shared knowledge of development and operations, as well as being co-located, meant that communications between the developers and operations was more natural and richer. The ITL describes how this resulted in fewer tickets being raised because "*you don't need a ticket, you go work within the team, … you have natural communication with the people around you and it's quite different. It's a big enabler when you can communicate naturally, I think*" [ITL]. He goes on to describe how the increased face-to-face communications (rather than email) between Dev and Ops also was a benefit in clarifying a misunderstanding: "*…within a couple of minutes you've resolved or clarified something that you would have spent, maybe 15 minutes to half an hour in trying to write out an email response.*"

## 6.5 Challenges in Adopting DevOps

During the year-long journey of DevOps implementation a number of challenges were identified by interviewees. These are aspects of implementing DevOps that slowed down the implementation by inhibiting enablers of DevOps or increasing the risk of not achieving the goals of DevOps. Figure 6 summarises the main areas of challenge (rectangular borders) and related issues. The lines depict hypothesized relationships of influence.

*6.5.1 Having staff with the right technical skills.* This challenge relates to both recruiting new staff with the technical skills as well as up-skilling and retaining current staff. The lack of appropriately skilled staff can lead to slowing down of the DevOps adoption journey because the capabilities needed are missing at the time of need. As discussed in section 6.3 in more detail, the skills relate to competency in writing software as well as understanding infrastructure and its setup, deployment, post-deployment monitoring, infrastructure problem solving, and skills in using the supporting tools.

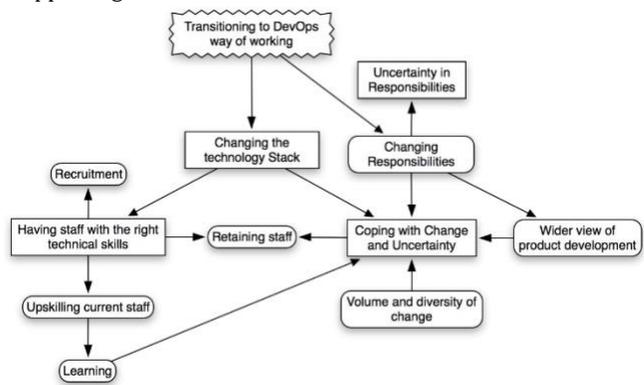

**Figure 6 Challenges related to DevOps Adoption**

The RQL viewed "*staffing as probably our biggest challenge*" and that there is a shortage of suitable job seekers and graduates because, in the opinion of the infrastructure team lead "*the skills set doesn't exist*". The Training manager emphasised the challenge of upskilling the entire team so anyone has the capability to be on call for operational problems. He described the upskilling of existing staff on the use of the new monitoring and automation tools and principles as currently a "bottleneck" to growth in DevOps adoption. From the team's perspective the challenge is the steep learning curve. As one Tester stated, the challenge is "*just keeping up because there are so many new tools and ideas*". One Developer also noted that, although the developers are used to learning emerging new technologies frequently, the challenge is to get enough high quality training to learn the Ops related technologies and ideas quickly enough to keep up with work demands.

*6.5.2 Resistance to Change and Uncertainty.* The transition to a DevOps way of working needs some motivation to overcome resistance to this long-term change and effort, and cope with the uncertainty of how this change will impact them in the future. As one Developer stated: "*I thought I was just going to write code*" and that DevOps "*was not what I signed up for*". The infrastructure team lead notes that it is a slow process getting the infrastructure experts to be accepted as part of the team and work effectively, as well as share knowledge. He states that "*you can't just slam them together and expect them to work because you've got two different skill sets and cultures initially*." He goes on to observe that acceptance of the change in mind set related to requiring *all* team members to be rostered as on-call for dealing with operational issues that arise was particularly challenging. The QA release manager had a view that the sheer volume and diversity of change related to the transition to DevOps is challenging for teams. She noted that changes may be needed in parallel and may be held up because of lack of resources or dependencies. She also observed that "*having so many balls in the air*" related to change can lead to disagreements or burnout. So resistance to change and uncertainty can slow down the availability of skilled staff through staff turnover from burnout or and slow upskilling, as well as slow acceptance of adoption of DevOps practices.

*6.5.3 Changing the Technology Stack and Tools.* The transition of the product to the cloud and a micro-services architecture was seen as a strong enabler of the adoption of DevOps and continuous deployment (as well as for other strategic business reasons). A year into the product re-architecting, the infrastructure team lead describes this part of the DevOps journey as having been incredibly complex and challenging. Similarly, deciding on, experimenting with, and setting up the tools for the build pipeline including full-stack testing, as well as automated deployment and monitoring has been challenging, according to an embedded Ops team member. He describes it as time-consuming, slow and technically complex, with "*no time for complacency*". The challenge of changing the technology stack is related to the challenge of finding the skilled staff to set and use the new technology stack, as well as the challenge of rapid learning and coping with this change and the associated uncertainty.

*6.5.4 Uncertainty in Responsibilities.* The shift in responsibilities associated with adopting DevOps is gradual and this has sometimes led to misunderstandings about who is responsible for what work activities. For example, the Tester describes the situation where ownership of infrastructure health is "*shifting but not fully shifted yet*", and this has led to misunderstanding: "*Sometimes I think you have taken care of such part, this part, but the other team think, okay, product team already take care of this bit [and it is missed]*".

# 7 DISCUSSION

Overall, the findings presented in the previous section align well with findings from other research in DevOps and provide more empirical support for this body of knowledge. In this section, the findings are compared with literature and implications for educators, practitioners and researchers discussed.

## 7.1 Meaning of DevOps

It is useful to have a consistent and clear understanding of the meaning of the term "DevOps" within an organization. If the meaning is not shared within an organisation, this increases the risk of misunderstandings, goal misalignment, and missed benefits. We found that the conceptualization DevOps was quite consistent and well developed in the organisation and aligned well with other researchers' findings. While Smeds & colleagues define DevOps as *enabling capabilities*, supported by *cultural* and *technological* enablers [7], others argue that the perceived meanings of DevOps depends on whether the emphasis is on the underlying goal for adopting DevOps or on the processes and practices through which collaboration between development and operations is achieved [26]. Findings from this study bridges these two views: while the main goal for adopting DevOps was to achieve continuous deployment of quality software, DevOps was seen as a way of integrating the processes, practices, roles and skill sets of development and operations closer together to align the incentives of the key personnel/roles (development, operations, and testing) involved in delivering software [12, 21]. Team traits and behaviors such as team ownership and team responsibility, and a number of technological tools and practices relating to automation, monitoring, and deployment were also important to the shared meaning of DevOps capabilities central to the meaning of DevOps in [7].

## 7.2 Drivers and realised benefits

In the organisation it was a business driver related to overcoming the limitations and frustrations of the current situation as well as the need to enhance the company's agility and competitive advantage that initiated the change to DevOps. This then translated into team drivers related improving speed, quality and release frequency. In agreement with the findings from the literature, the case organisation experienced some expected benefits of DevOps adoption such as increased frequency of quality deployments, improved quality assurance, and increased collaboration between development and operation teams [8]. In addition, the findings also reveal the influence of some key relationships between the realized benefits. For example, benefits such as high autonomy, learning new technologies, feeling valued, and motivating collaboration contributed to improved team morale and engagement. And while benefits such as improved code quality, natural communications, and knowledge sharing contribute positively to improved deployment frequency, the benefits of improved frequency of releases and improved application quality in turn contribute to improving customer experience.

### 7.3 Enablers

The technical and capability enablers aligned well with those suggested by Smeds and colleagues [7]. Furthermore, they were generally implemented to a reasonable level. The architecture switch to cloud-based delivery and micro-services was also seen as an important technical enabler by the interviewees. The additional technical enabler of automated measurement and capability enabler of continuous measurement were not implemented to at this stage of the DevOps adoption case, at the time of the interviews. This was reflected in the low visibility and qualitative nature of the benefits accrued from adopting DevOps.

One clear theme from the interviews was that the adoption of the technical enablers that supported the capability enablers was a complex, gradual process taking considerable effort and resources. This is reflected in some of the technical challenges case identified by the interviewees.

### 7.4 Challenges

A number of main challenges in adopting DevOps have been identified in the literature: lack of clear definition[7, 8], insufficient communication [8], deep-seated company culture [8], organisation structure, and geographical distribution [7]. However, not all of those challenges were evidenced in the findings of this case study. For example, the lack of clear definition was not perceived as a major challenge, as there was a company-centric understanding (*embedded ops)* and consensus about the meaning of DevOps. Challenges related to geographic distribution was not applicable as the development and operations work in the company was not distributed. However, there were other aspects that either slowed down the implementation by inhibiting enablers of DevOps or increasing the risk of not achieving the goals of DevOps. For example, interviewees highlighted challenges related to (i) recruiting new staff with appropriate technical skills and training, (ii) providing high quality training to existing staff, and (iii) retaining current staff who had the relevant qualification, skills and experience. Other challenges associated with shifting responsibilities and the volume and diversity of change related to the transition to DevOps were perceived to create resistance to change and uncertainty. Provisioning appropriate technologies and tools such as cloud hosting platform, a micro-services architecture, and experimenting with automated deployment and monitoring was also perceived as extremely complex and challenging. Similar to the relationships between realized benefits, the findings also highlight the influence of relationships between the challenges. For example, the challenge of changing the technology stack is related to the challenge of finding the skilled staff, as well as the challenge of coping with change and the associated uncertainty.

## 8 THREATS TO VALIDITY

There are likely to be researchers' biases influencing the interpretations of the qualitative analyses of the interview data. To reduce this bias, the data analysis was collaborative and the results were discussed between two researchers until consensus was reached. The high-level categorisations were also reviewed by a member of the case organisation.

Construct validity relates to ensuring there is a shared understanding of the language and terminology among the interviewees and other researchers so that the interview questions were interpreted in the manner intended. One researcher conducted all the interviews using the same interview guide for each interview. This consistent interview protocol included explaining the purpose of the survey, inviting clarification questions at any stage, and explaining the main terminology. Prior to the interviews, the interview questions were reviewed for ambiguities and biases by the researchers and a pilot interview was conducted with an expert from industry.

To avoid leading the interviewees to answers or guessing expected conclusions, the interviewer retained a neutral stance about interviewees' explanations and descriptions. The second author (and interviewer) has had involvement with the case organisation for several years and so there was already a basis for mutual trust. This could lower the likelihood of the interviewees being influenced by the presence of the researcher.

The external validity of case studies is generally low because of the uncertain effects of changing contextual variables such as project and team characteristics. The findings from our single case study could hardly be claimed to be generalizable to other contexts. The qualitative findings can be considered as hypotheses, rather than facts that are valid in general, and form the basis of future research.

## 9 CONCLUSION

Our study presents findings of an in-depth exploratory case study that investigated DevOps implementation in a New Zealand product development organisation. Our investigation explored the meaning of DevOps, the main drivers, enablers, and benefits and challenges of adopting DevOps. For the case organisation, DevOps was "embedded ops", which implied optimal team combinations in which operations could be embedded within a team of developers and testers or spread across a few teams. The meaning of DevOps as expressed by the interviewees, was seen as a way of integrating the roles and skill sets of development and operations closer together to align the incentives of the key roles involved in delivering software. The support of team traits and behaviors such as team ownership and team responsibility, and technological enablers such as implementing an automation pipeline and cross functional organisational structures, were critical to delivering the expected benefits of DevOps.

The realized benefits of DevOps adoption included increased frequency of quality deployments, and increased collaboration between development and operation teams. The influence of key relationships between the realized benefits was identified. For example, while benefits such as high autonomy, motivating collaboration, and feeling valued contributed to improved team morale and engagement, benefits such as improved code quality, natural communications, and knowledge sharing were found to contribute positively to improved deployment frequency.

The case organisation experienced a number of challenges that slowed down the DevOps implementation process. These included challenges related to recruiting new staff with appropriate technical skills and training, providing high quality training to existing staff, and retaining current staff who had the relevant qualifications, skills and experience. Challenges associated with shifting responsibilities and the volume and diversity of change created some resistance to change and uncertainty. Provisioning appropriate technologies and tools such as cloud hosting platform, a micro-services architecture, and experimenting with automated deployment and monitoring were also identified as challenges.